\documentclass[12pt]{iopart}
\usepackage{bm,amssymb,graphicx}
\usepackage{iopams}

\begin{document}

\title[The Fourier state of dilute granular gases]{The Fourier state of a dilute granular gas described by the inelastic Boltzmann equation}

\author{J. Javier Brey, N. Khalil, and M.J. Ruiz-Montero}

\address{F\'{\i}sica Te\'orica, Universidad de Sevilla, Apartado
de Correos 1065, E-41080 Sevilla, Spain} \ead{brey@us.es}

\begin{abstract}
The existence of two stationary solutions of the nonlinear Boltzmann equation for inelastic hard spheres or disks is investigated. They are restricted neither to weak dissipation nor to small gradients. The one-particle distribution functions are assumed to have an scaling property, namely that all the position dependence occurs through the density and the temperature. At the macroscopic level, the state corresponding to both is characterized by  uniform pressure, no mass flow, and a linear temperature profile. Moreover, the state exhibits two peculiar features. First, there is a relationship between the inelasticity of collisions, the pressure, and the
temperature gradient. Second, the heat flux can be expressed as being linear in the temperature gradient, i.e. a Fourier-like law is obeyed. One of the solutions is singular in the elastic limit. The theoretical predictions following from the other one are compared with molecular dynamics simulation results and a good agreement is obtained in the parameter region in which the Fourier state can be actually observed in the simulations, namely not too strong inelasticity.

\end{abstract}



\maketitle

\section{Introduction}
\label{s1} The knowledge of dilute granular gases has
improved greatly in the last years \cite{PyB03,ByP04,Go03}. Among many
others, one of the fundamental reasons for studying granular gases
is that they are considered as a proving ground for kinetic theory
and non-equilibrium statistical mechanics. The underlying idea is
that the grains can be assimilated to atoms or molecules and,
therefore, granular gases to molecular gases. Due to the inherent
energy dissipation in collisions between grains, there is no
equilibrium state for granular gases and any state of them is of a
non-equilibrium nature. At a phenomenological level, granular gases
exhibit many similarities with ordinary gases, although also some
strong differences \cite{AyT06}.

Grains are often modeled as inelastic hard spheres or disks and, in
the simplest versions, tangential friction is neglected. For this
kind of models, the inelastic Boltzmann kinetic equation
\cite{GyS95} provides the accurate starting point  to study
low-density granular gases. The validity of this equation is
justified on exactly the same grounds as its elastic, molecular,
limit. It can be derived either heuristically or from the Liouville
equation in the asymptotic small density limit \cite{BDyS97}.
Moreover, the theoretical predictions derived from it have been
found to be in very good agreement with molecular dynamics simulation
results, when the comparison is carried out inside the appropriate
range of the parameters defining the system. The above includes the
velocity distribution of the simplest possible state of a granular
gas, the so-called homogeneous cooling state \cite{vNyE98}, and its
exponential tail \cite{EyP97,HOyB00}, the inelastic hydrodynamic
equations to Navier-Stokes order, and the expressions for the
transport coefficients appearing in them \cite{ByC98,BRCyG00}. More
severe conditions such as the initial departure of a granular gas
from homogeneity due to an instability have also been investigated
\cite{ByR04}, finding again that the Boltzmann equation accurately
predicts the behavior of the system in the low density limit.

The Chapman-Enskog procedure for deriving hydrodynamic equations,
assumes the existence of a special
kind of solutions of the Boltzmann equation, the normal solutions,
and that they can be constructed by means of an expansion in a
formal inhomogeneity parameter \cite{RydL77,Mc89}. The hydrodynamic
Navier-Stokes equations correspond to keeping up to the first order in
the gradients contributions to the distribution function, giving
rise to second order in the gradients terms in the hydrodynamic
equations, when employed in the formal expressions of the heat and
momentum fluxes. A peculiarity of granular gases is that
hydrodynamic gradients are often induced by inelasticity. This
holds particularly in the case of steady states, since the only way
of compensating for the energy dissipation in collisions is through
energy fluxes associated to gradients in the system. Consequently,
in these systems there is a coupling between gradients and
inelasticity, so that the restriction to small gradients, as it is
the case of the Navier-Stokes equations, implies in many cases the
restriction to small inelasticity too. For some states, like the
steady shear flow, the situation is even more complex, since they
are inherently rheological and the Navier-Stokes approximation never
applies \cite{SGyD04}.

Of course, both limitations mentioned above, the assumption of a
normal solution and of an expansion in powers of the non-uniformity
of the system, are overcome if the  exact solution of the Boltzmann
equation for a given state is known. In this case, all the transport
properties can be computed with no assumption and/or approximation.
The problem is that very little is known about exact inhomogeneous
solutions of the Boltzmann equation that are relevant for transport.
For molecular, elastic systems, the available exact solutions
correspond to the very unrealistic case of Maxwell molecules
\cite{IyT56,AMyN80}. On the other hand, a few years ago the
existence of an inhomogeneous exact solution of the Boltzmann
equation for smooth inelastic hard spheres or disks was suggested
\cite{BCMyR01}. It has the property that all the spatial dependence
in the velocity distribution occurs through its second velocity
moment or, equivalently, the granular temperature. The macroscopic
state is stationary and with gradients in only one direction.
Moreover, the relationship between the heat flow and the temperature
gradient can always be expressed in a  linear form, no matter the value of the parameters of the
system. For this reason, it was referred to as the Fourier state.

Here the study of the Fourier state is undertaken in more detail.
At the beginning, the aim of this study was twofold: to construct a solution of the equations
derived in \cite{BCMyR01} beyond the Navier-Stokes approximation
considered there, and to investigate at what extent the predicted
state can be actually observed in molecular dynamics simulations.
The latter is a first step towards the possibility of seeing the
state in real experiments. Nevertheless, the analysis to be presented
indicates the possible existence of two different solutions of the
inelastic Boltzmann equation having the properties associated to the
Fourier state as described above. The first of them agree, in the appropriate
range of parameters, with the solution obtained in the Navier-Stokes approximation
following from the Chapman-Enskog solution to the (inelastic) Boltzmann equation.
On the other hand, the other solution can not be inferred from a Chapman-Enskog-like
algorithm. The reason is that it is singular, in the sense that its elastic limit does not
correspond to any solution of the elastic Boltzmann equation. For similar reason, this solution is not
captured either by the hydrodynamic Navier-Stokes equations for dilute granular gases.

The remainder of the paper is organized as follows. The Fourier state
of a dilute granular gas is defined in Sec. \ref{s2}, where the assumed form of the
one-particle distribution function for this state is reviewed. Substitution
of its expression into the inelastic Boltzmann equation and scaling of the velocity,
leads to a closed kinetic equation in which all the explicit space dependence has been eliminated.
In Sec. \ref{s3}  a Sonine expansion of the distribution function solution of the Boltzmann
equation is considered. The expansion is truncated to the lowest order leading to nontrivial
contributions to the even and odd parts of the distribution function. Equations for
the coefficients of the remaining terms are obtained by taking velocity moments in the Boltzmann equation.
In these equations only the lowest nonlinear contributions are kept. Then, two different solutions, referred
to as  ``regular'' and ``singular'', respectively, are identified. The origin and consistency
of the singular solution is discussed in Sec.\  \ref{s4}. The purpose there is not to prove rigourously the
existence of the solution, but to check the consistency of the calculations and approximations
indicating that this is the case.

A comparison of the theoretical predictions with molecular dynamics simulation results is
carried out in Sec.\  \ref{s5}. The state described by the singular solution of the Boltzmann equation
has never been observed. A possible and probable reason for it is that this solution corresponds to
a highly unstable state. On the other hand, for weak inelasticity, a good agreement is found in the bulk of the
simulated systems with the predictions from the regular solution. As the inelasticity is increased, the Sonine approximation fails and
later on, the Fourier state becomes very difficult to reach, in a significant space region of the system, in the simulations.
Section \ref{s6} presents a brief discussion of the results, possible extensions, and also some open questions.

\section{The Fourier state}
\label{s2} Consider a dilute granular gas composed of smooth
inelastic hard spheres ($d=3$)  or disks ($d=2$) of diameter
$\sigma$ and mass $m$, being $\alpha$ the velocity-independent
coefficient of normal restitution. The one-particle distribution
function $f({\bm r},{\bm v},t)$ is assumed to obey the inelastic
Boltzmann equation \cite{GyS95}. For  steady distributions depending
on position only through the $x$-coordinate, it has the form
\begin{equation}
\label{2.1} v_{x} \frac{\partial}{\partial x}\, f(x,{\bm v})= J[{\bm
v}|f],
\end{equation}
where the Boltzmann collision term $J$ is a functional of the form
\begin{eqnarray}
\label{2.2} J[{\bm v}|f] & \equiv & \sigma^{d-1} \int d{\bm v}_{1}\ \int
d \widehat{\bm \sigma}\, \theta ({\bm g} \cdot \widehat{\bm \sigma})
{\bm g} \cdot \widehat{\bm \sigma} \left[ \alpha^{-2} f(x,{\bm
v}^{\prime}) f(x,{\bm v}^{\prime}_{1}) \right. \nonumber \\
&& \left. - f(x,{\bm v}) f(x,{\bm
v}_{1}) \right].
\end{eqnarray}
In the above expression, $g \equiv {\bm v}-{\bm v}_{1}$ is the
relative velocity, $\widehat{\bm \sigma}$ is a unit vector along the
line of centers of the two colliding particles at contact, $\theta$
is the Heaviside step function, and ${\bm v}^{\prime}$, ${\bm
v}^{\prime}_{1}$ are the precollisional velocities leading to ${\bm
v}$ and ${\bm v}_{1}$. Moreover, attention here will be restricted
to solutions having a vanishing average velocity, i.e.,
\begin{equation}
\label{2.2a} \int d{\bm v}\, {\bm v} f(x,{\bm v})= 0.
\end{equation}
Balance equations are obtained in the usual way  by taking velocity
moments in Eq. (\ref{2.1}),
\begin{equation}
\label{2.3} \frac{\partial }{\partial x}\, P_{x, i}=0,
\end{equation}
\begin{equation}
\label{2.4} 2(nd )^{-1} \frac{\partial q_{x}}{\partial x}+T(x) \zeta
(x)=0,
\end{equation}
where $n(x)$ and $T(x)$ are the number density and the temperature,
respectively,
\begin{equation}
\label{2.5} n(x) \equiv \int d{\bm v}\, f(x,{\bm v}), \quad \frac{d
n(x)T(x)}{2} \equiv \int d{\bm v} \frac{mv^{2}}{2}\, f(x,{\bm v}),
\end{equation}
and the pressure tensor $P_{ij}$ and heat flux ${\bm q}$ are defined
as
\begin{equation}
\label{2.6} P_{ij} \equiv \int d{\bm v}\, m v_{i}v_{j} f(x,{\bm v}), \quad
{\bm q} \equiv \int d{\bm v}\, \frac{m v^{2}}{2}\, {\bm v} f(x,{\bm
v}).
\end{equation}
Finally, $\zeta(x)$ is the cooling rate due to the energy
dissipation in collisions. Its expression is
\begin{equation}
\label{2.7} \zeta(x)=\frac{m \pi^{\frac{d-1}{2}} \sigma^{d-1}
(1-\alpha^{2})}{4d \Gamma \left( \frac{d+3}{2} \right) nT}
 \int d{\bf v}
\int d{\bf v}_{1}\, g^{3} f(x,{\bf v}) f(x,{\bf v}_{1}).
\end{equation}

Solving Eq.\ (\ref{2.1}) for given boundary conditions, is a very
hard task. In ref. \cite{BCMyR01}, the search of a solution having
the scaling form
\begin{equation}
\label{2.8} f(x,{\bf v})=n(x) \left[ \frac{m}{2 T(x)} \right]^{d/2}
\varphi ({\bm c}), \quad {\bm c} \equiv \left[ \frac{m}{2T(x)}
\right]^{1/2} {\bf v},
\end{equation}
was proposed. This expression is similar to the distribution
function of the so-called homogeneous cooling state \cite{GyS95},
but with the position $x$ playing the role of the time. The above
functional form had already been used in ref. \cite{GZyB97} to fit the molecular
dynamics results for the distribution function of the steady state of
a granular gas with a temperature gradient.
Equations (\ref{2.3}) and  (\ref{2.8}) imply that the pressure tensor and,
therefore, the pressure $p=\sum_{i} P_{ii}/d=nT$ are uniform. Due to Eqs. (\ref{2.3})
and (\ref{2.5}), the function $\varphi$ must verify
\begin{equation}
\label{2.9} \int d{\bm c}\, \varphi ({\bm c})=1,\quad \int d{\bm
c}\, {\bm c} \varphi ({\bm c})=0, \quad \int d{\bm c}\, c^{2}
\varphi ({\bm c})= \frac{d}{2}.
\end{equation}
Use of Eq.\ (\ref{2.8}) into Eq.\ (\ref{2.4}), taking into account Eqs.\
(\ref{2.6}) and (\ref{2.7}), yields
\begin{equation}
\label{2.10} \frac{dT}{dx}= p \sigma^{d-1} I[\varphi]
\end{equation}
with
\begin{equation}
\label{2.11} I[\varphi] \equiv - \frac{(1-\alpha^{2})
\pi^{\frac{d-1}{2}}}{2 \Gamma \left( \frac{d+3}{2} \right)}\,
\frac{\int d{\bm c}\int d{\bm c}_{1}\, |{\bm c}-{\bm c}_{1}|^{3}
\varphi ({\bm c}) \varphi ({\bm c}_{1})}{\int d {\bm c}\, c^{2}
c_{x} \varphi ({\bm c})}\, .
\end{equation}
It follows that the temperature profile is strictly linear in $x$,
\begin{equation}
\label{2.12} \frac{dT}{dx} = \theta = {\mbox constant}.
\end{equation}
Moreover, Eq. (\ref{2.10}) establishes a relationship between the
pressure, the temperature gradient, and the coefficient of
restitution $\alpha$ (through the function $\varphi$),
\begin{equation}
\label{2.13} \frac{\theta}{p \sigma^{d-1}} = I[\varphi].
\end{equation}
The heat flux in the $x$ direction is
\begin{equation}
\label{2.14} q_{x} = \left[\frac{2 T(x)}{m} \right]^{1/2} p
\int d{\bm c}\,  c^{2}c_{x} \varphi ({\bm c})= \left( \frac{2T}{m}
\right)^{1/2} \frac{\int d {\bm c}\,  c^{2} c_{x} \varphi ({\bm
c})}{\sigma^{d-1} I[\varphi]}\, \frac{dT}{dx}.
\end{equation}
This has the form of a Fourier law, with the heat flux coupled
linearly to the temperature gradient. It is worth to stress that
here it has been derived without any explicit restriction to small
gradients. This is the reason why this state was referred to as the
Fourier state \cite{BCMyR01}.

To get a closed equation for the function $\varphi$, Eq. (\ref{2.8}) is substituted
into the Boltzmann equation (\ref{2.1}), taking into account Eq.\
(\ref{2.10}). The result reads:
\begin{equation}
\label{2.15} -I[\varphi]\left\{ c_{x} \varphi ({\bm c})+ \frac{c_{x}}{2}
\frac{\partial}{\partial {\bm c}}\, \cdot \left[ {\bm c}
\varphi ({\bm c}) \right] \right\}= \sigma^{1-d} J[{\bf c}|\varphi].
\end{equation}
From a mathematical point of view, the problem is fully analogous to the
identification of the distribution function of the homogeneous
cooling state \cite{GyS95}.

\section{Sonine expansion}
\label{s3}
Because of symmetry, the function $\varphi({\bm c})$ must be an even function of
the vector component of ${\bm c}$ perpendicular to the $x$-axis, ${\bm c}_{\perp}$.
Then, $\varphi$ is expanded in series of Sonine polynomials as
\begin{eqnarray}
\label{3.1}
\varphi({\bm c}) &  = & \pi^{-d/2} e^{-c^{2}} \sum_{i=0}^{\infty} \sum_{j=0}^{\infty}
\left[a_{ij} S_{-1/2}^{(i)} (c_{x}^{2})S_{\frac{d-3}{2}}^{(j)} (c_{\perp}^{2}) \right. \nonumber \\
&& \left. + b_{ij} c_{x}
S_{1/2}^{(i)} (c_{x}^{2})S_{\frac{d-3}{2}}^{(j)}(c_{\perp}^{2}) \right].
\end{eqnarray}
Here, the function has been decomposed into its even and odd in $c_{x}$ parts, and the
$S_{n}^{(i)}$ are the Sonine polynomials \cite{RydL77,Mc89},
\begin{equation}
\label{3.2}
S_{n}^{(i)}(z) \equiv \sum_{k=0}^{i} \frac{(-1)^{k} \Gamma (n+i+1)}{\Gamma (n+k+1)(i-k)! k!}\, z^{k}.
\end{equation}
They verify the orthogonality relation
\begin{equation}
\label{3.3}
\int_{0}^{\infty} dz\, z^{n} e^{-z} S_{n}^{(i)} (z) S_{n}^{(j)}(z) = \frac{\Gamma (n+i+1)}{i!}\,
\delta_{ij},
\end{equation}
where $\Gamma (z)$ is the Gamma function and $\delta_{ij}$ the Kronecker delta. In particular, it is
\begin{equation}
\label{3.4}
S_{n}^{(0)}(z)=1, \quad S_{n}^{(1)}(z) = n+1-z,
\end{equation}
for all $n$. Using the relation (\ref{3.3}), it is obtained that the necessary and sufficient conditions
for $\varphi$ to verify Eqs. (\ref{2.9}) are
\begin{equation}
\label{3.5}
a_{00}=1, \quad b_{00}=0, \quad a_{10}+(d-1)a_{01}=0.
\end{equation}
In the following, the expansion given in Eq.\ (\ref{3.1}) will be considered to the lowest nontrivial
order, keeping only the terms proportional to $a_{01}$ (and $a_{10}$), $b_{01}$, and $b_{10}$, and neglecting the remaining terms. This corresponds to the so-called first Sonine approximation
\cite{Mc89}. Explicitly, the approximation considered can be expressed as
\begin{eqnarray}
\label{3.5a}
\varphi ({\bm c}) & = & \pi^{-d/2} e^{-c^{2}} \Big[ 1- a_{01} (c^{2}-dc_{x}^{2}) + \left( \frac{d-1}{2}\, b_{01} + \frac{3}{2}\, b_{10} \right)c_{x}   \nonumber \\
& &  -b_{01} c_{x}c^{2} - (b_{10}-b_{01}) c_{x}^{3} \Big].
\end{eqnarray}
It is worth to emphasize that this approximation does not imply by itself the isotropy of the state. For instance, it is
\begin{equation}
\label{3.6}
\int d{\bm c}\ c_{x}^{2} \varphi ({\bm c}) = \frac{1}{2} + \frac{(d-1)a_{01}}{2}, \quad \int d{\bm c}\ c_{\perp}^{2} \varphi ({\bm c}) =\frac{d-1}{2}\, \left( 1-a_{01} \right),
\end{equation}
indicating that the diagonal elements of the pressure tensor are not the
same if $a_{01}$ does not vanish. Also, the dimensionless heat flux
\begin{equation}
\label{3.7}
Q_{x} \equiv  \left[ \frac{m}{2T(x)} \right]^{1/2} \frac{q_{x}}{p} = \int d{\bm c}\ c^{2} c_{x} \varphi ({\bm c}),
\end{equation}
can be decomposed into the two components
\begin{equation}
\label{3.8}
Q_{xx} \equiv \int d{\bm c}\,  c_{x}^{3} \varphi ({\bm c}) = - \frac{3 b_{10}}{4},
\end{equation}
\begin{equation}
\label{3.9}
Q_{x \perp} \equiv \int d{\bm c}\, c_{x} c_{\perp}^{2}   \varphi ({\bm c})= - \frac{(d-1)b_{01}}{4}.
\end{equation}
This implies the anisotropy in the energy being carried in the $x$ direction.
To identify the coefficients $a_{01}$, $b_{01}$, and $b_{10}$, velocity moments are taken in Eq.
(\ref{2.15}), using the expansion of $\varphi ({\bm c})$ in the first Sonine approximation. The equation
contains terms that are nonlinear in $\varphi ({\bm c})$, leading consequently to contributions that are nonlinear in the coefficients to be determined. Here the simplification is made of keeping only terms up to second degree in the coefficients $a_{01}$, $b_{01}$, and $b_{10}$, i.e. neglecting those terms proportional to
$a_{01}^{n_{1}} b_{01}^{n_{2}} b_{10}^{n_{3}}$ with $n_{1}+n_{2}+n_{3} \geq 3$.
A similar approximation to determine the coefficient characterizing the first Sonine approximation to the distribution function of the homogeneous cooling state \cite{GyS95,vNyE98}, was found to lead to a very good estimate, even for strong inelasticity
\cite{ByP04,ByP00}. In the present case, the accuracy of the approximation being used is just assumed for the sake of simplicity, without further justification.

Consider $I[\varphi]$ defined in Eq.\ (\ref{2.11}). A quadratic calculation as indicated above
gives
\begin{equation}
\label{3.10}
\int d{\bm c} \int d{\bm c}_{1}\, |{\bm c}-{\bm c}_{1} |^{3} \varphi ({\bm c}) \varphi ({\bm c}_{1})
= \frac{ 2^{3/2} \Gamma \left( \frac{d+3}{2} \right)}{\Gamma (d/2)}  \left[ 1+ g(b_{01},b_{10}) \right],
\end{equation}
\begin{equation}
\label{3.11}
g(b_{01},b_{10}) \equiv  \frac{45 b_{10}^{2}+ 18(d-1)
b_{01}b_{10}+3(d^{2}-1)b_{01}^{2}}{64d(d+2)(d+4)}
\end{equation}
and then, using Eqs. (\ref{3.7})-(\ref{3.9}), it follows that
\begin{equation}
\label{3.12}
I[\varphi]= \frac{(1-\alpha^{2}) \pi^{\frac{d-1}{2}} 2^{5/2}}{\Gamma(d/2) \left[ (d-1)b_{01} + 3 b_{10} \right]} \left[ 1+g(b_{01},b_{10})\right].
\end{equation}
Now, Eq.\ (\ref{2.15}) is multiplied by $c_{x}^{2}$ and integrated over ${\bm c}$. After some algebra, it is obtained
\begin{eqnarray}
\label{3.13}
(1-\alpha)(d-1) (b_{01}-3b_{10}) & = & - a_{01} \frac{(d-1)(2d+3-3 \alpha)}{(d+2)}\, \nonumber \\
& & \times \left[ (d-1)b_{01}+ 3b_{10} \right].
\end{eqnarray}
Upon obtaining this expression,  Eq.\ (\ref{3.11}) has been employed. Notice that an independent relationship between the coefficients can not be derived by multiplying Eq. (\ref{2.15}) by $c_{\perp}^{2}$ and posterior integration over ${\bm c}$. The reason is that the balance equation for the energy has already been employed in the derivation of Eq.\ (\ref{2.15}). Therefore, in order to get two more independent equations for the coefficients $a_{01}$, $b_{01}$, and $b_{10}$, higher velocity moments of Eq.\ (\ref{2.10}) have been considered. In the Appendix, the equations obtained by multiplying the kinetic equation by $c_{x}^{3}$ and by $c^{2}c_{x}$, and integrating afterwards over the velocity ${\bm c}$, are reported. Then the parameters $a_{01}$, $b_{01}$, and $b_{10}$ can be obtained by numerically solving
the system formed by Eqs.\ (\ref{3.13}), (\ref{ap.1}), and (\ref{ap.2}). The equations are of second degree and have two different sets of real solutions. They will be referred to as ``regular'' solution and ``singular'' solution, respectively. The reason for this nomenclature will be discussed later on.

\begin{figure}
\begin{center}
\includegraphics[scale=0.7,angle=0]{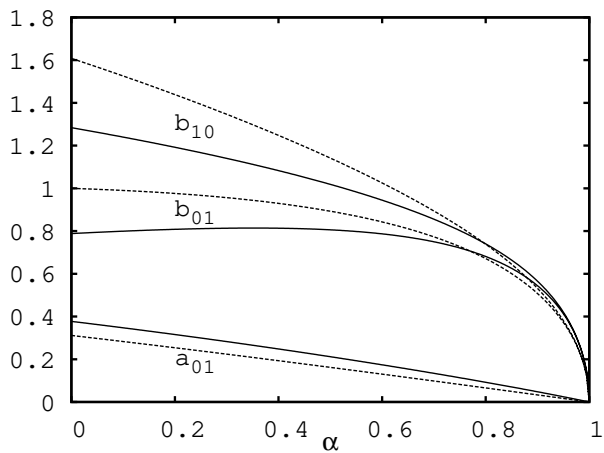}
\caption{The dimensionless parameters $a_{01}$, $b_{10}$, and $b_{01}$ determining the distribution function of the ``regular'' Fourier state in the first Sonine approximation considered in the text, as a function of the coefficient of normal restitution $\alpha$. The solid lines correspond to $d=2$ and the dashed ones to $d=3$. \label{fig1}}
\end{center}
\end{figure}

\begin{figure}
\begin{center}
\includegraphics[scale=0.7,angle=0]{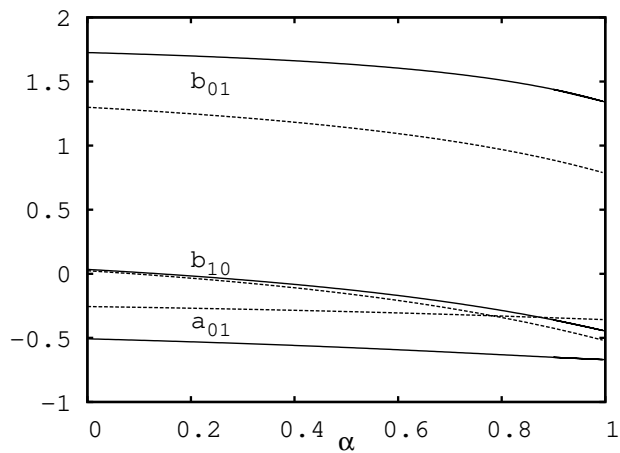}
\caption{The dimensionless parameters $a_{01}$, $b_{10}$, and $b_{01}$ determining the distribution function of the ``singular'' Fourier state in the first Sonine approximation considered in the text, as a function of the coefficient of normal restitution $\alpha$. The solid lines correspond to $d=2$ and the dashed ones to $d=3$. \label{fig2}}
\end{center}
\end{figure}

In Fig. \ref{fig1} the parameters $a_{01}$, $b_{01}$, and $b_{10}$ are plotted as a function of $\alpha$ for the regular solution. The results corresponding to both $d=2$ and $d=3$ are shown. The parameters defining the singular solution are given in Fig.\ \ref{fig2}.  It is seen that $a_{01}$ does not vanish in any of the two solutions for $\alpha \neq 1$, then indicating the anisotropy of the pressure tensor, as discussed above. In the limit $\alpha \rightarrow 1$, there is a strong difference between the behaviors of the regular and the singular solutions. While in the former the three coefficients tend to vanish, in the latter they tend to finite, non-vanishing values. Moreover, for the regular solution $a_{01}$ becomes very small and $b_{01} \simeq b_{10}$, in the limit of small inelasticity ($\alpha$ close to one). This is consistent with the results obtained to
Navier-Stokes order by the Chapman-Enskog procedure in the first Sonine approximation, that leads to a distribution function verifying  $a_{01}=0$, $b_{01}=b_{10}$ \cite{BDKyS98}.  When the inelasticity of the system increases, the values of $b_{01}$ and $b_{10}$ for the regular solution grow both very fast and the difference between them becomes significant. On the other hand, the behavior of the singular solution for $\alpha \rightarrow 1$ strongly differs from the Chapman-Enskog solution, since not only the three coefficients tend to non-zero values, but the limits of $b_{01}$ and $b_{10}$ are definitely different, having even opposite sign. A direct consequence  of this behavior is that, while the first solution tends to the elastic equilibrium Gaussian for $\alpha \rightarrow 1$, the other one does not. This is the reason to qualify them as regular and singular solutions, respectively.

\section{The singular solution}
\label{s4}
\begin{figure}
\begin{center}
\includegraphics[scale=0.7,angle=0]{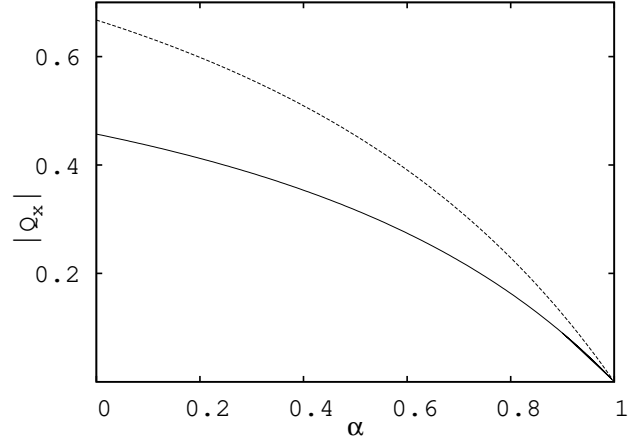}
\caption{The dimensionless heat flux $Q_{x}$ for the singular solution of the inelastic Boltzmann equation for the Fourier state. The solid line correspond to $d=2$ and the dashed one to $d=3$. \label{fig3}}
\end{center}
\end{figure}

It has been mentioned above that a peculiarity of the singular solution of the Boltzmann equation for the Fourier state is that the coefficients $b_{01}$ and $b_{10}$ do not vanish in the elastic limit, tending to values with opposite signs in the elastic limit. A direct implication of this, following from Eqs.\ (\ref{3.8}) and (\ref{3.9}) is that the two components of the heat flux, $Q_{xx}$ and $Q_{x \perp}$, also have opposite signs. That means that particles moving in one direction have on the average larger values of $v_{x}^{2}$, while particles moving in the opposite direction have, also on the average, larger values of $v_{\perp}^{2}$. Although one can be prompted to conclude that something fundamental is being violated by this solution, we have not been able to identify any argument leading to discard it a priori. A trivial first test of the consistency of the singular solution is that the net heat flux must vanish in the elastic limit, since the dissipation and, therefore, the temperature gradient vanish in it.  In Fig. \ref{fig3}  the dimensionless heat flux $Q_{x}$ corresponding to the singular solution is plotted as a function of the coefficient of normal restitution. It is seen to vanish when $\alpha \rightarrow 1$, for both $d=2$ and $d=3$, as it should.

In order to get additional information about whether the singular solution is an artifact  of the Sonine approximation carried out in sec. \ref{s3}, it seems worth to investigate the asymptotic behavior of the solutions of the  Boltzmann equation
(\ref{2.15}) in the limit $\alpha \rightarrow 1$. Assume that in this limit, the asymptotic behavior of the solutions
of the equation reads
\begin{equation}
\label{4.1}
\varphi ({\bm  c}) \sim \varphi^{(0)}  ({\bm c}) + \epsilon^{q} \varphi^{(1)}  ({\bm c}),
\end{equation}
where $\epsilon \equiv (1- \alpha^{2})^{1/2}$, and the parameter $q>0$ and the functions $\varphi^{(0)}  ({\bm c})$ and $\varphi^{(1)}  ({\bm c})$ are to be determined in the following by consistency. As a consequence,
\begin{equation}
\label{4.2}
Q_{x} \sim Q_{x}^{(0)} + \epsilon^{q} Q_{x}^{(1)},
\end{equation}
with
\begin{equation}
\label{4.3}
Q_{x}^{(0),(1)} = \int d{\bm c}\, c^{2}c_{x} \varphi^{(0),(1)}  ({\bm c}).
\end{equation}
Similarly,
\begin{equation}
\label{4.4}
\gamma[\varphi] \equiv \frac{\pi^{\frac{d-1}{2}}}{2 \Gamma \left( \frac{d+3}{2} \right)}\,
\int d{\bm c}\int d{\bm c}_{1}\, |{\bm c}-{\bm c}_{1}|^{3}
\varphi ({\bm c}) \varphi ({\bm c}_{1})
\sim \gamma^{(0)} + \epsilon^{q} \gamma ^{(1)},
\end{equation}
\begin{equation}
\label{4.5}
\gamma^{(0)} =
\frac{\pi^{\frac{d-1}{2}}}{2 \Gamma \left( \frac{d+3}{2} \right)}\,
\int d{\bm c}\int d{\bm c}_{1}\, |{\bm c}-{\bm c}_{1}|^{3}
\varphi^{(0)}
 ({\bm c}) \varphi^{(0)} ({\bm c}_{1}),
\end{equation}
\begin{equation}
\label{4.6}
\gamma^{(1)} =
\frac{\pi^{\frac{d-1}{2}}}{ \Gamma \left( \frac{d+3}{2} \right)}\,
\int d{\bm c}\int d{\bm c}_{1}\, |{\bm c}-{\bm c}_{1}|^{3}
\varphi^{(0)}
 ({\bm c}) \varphi^{(1)} ({\bm c}_{1}).
\end{equation}
The asymptotic behavior of the inelastic Boltzmann collision operator is easily identified as
\begin{equation}
\label{4.7}
J \left[ {\bm c}|g \right] \sim J^{(0)} \left[ {\bm c}| g \right] + \epsilon^{2} J^{(2)} \left[ {\bm c}| g \right],
\end{equation}
where $J^{(0)}$ is the elastic Boltzmann collision operator and the explicit for of $ J^{(2)} \left[ {\bm c}| g \right]$ will not be needed in the following.

Putting $\epsilon =0$ in Eq.\ (\ref{2.15}) it is obtained
\begin{equation}
\label{4.8}
Q_{x}^{(0)} J^{(0)}[{\bm c} | \varphi^{(0)}] =0.
\end{equation}
Therefore, either
\begin{equation}
\label{4.9}
Q_{x}^{(0)}=0
\end{equation}
or
\begin{equation}
\label{4.10}
J^{(0)}[{\bm c} | \varphi^{(0)}] =0.
\end{equation}
Since Eq.\ (\ref{4.10}) implies that $\varphi^{(0)} ({\bm c})$ is the Gaussian for which the
dimensionless heat flux vanishes, it is concluded that Eq. (\ref{4.9}) is always verified. When also
Eq.\ (\ref{4.10}) is fulfilled, the behavior corresponding to the regular solution, and to the Chapman-Enskog expansion, is recovered. The other possibility, namely $J^{(0)}[{\bm c} | \varphi^{(0)}]  \neq 0$  but still $Q_{x}^{(0)}=0$, is expected to
lead to the singular solution. To investigate further this issue, the next order balance in the Boltzmann equation
(\ref{2.15}) is considered. It is given by
\begin{equation}
\label{4.11}
\epsilon^{2} c_{x}  \gamma^{(0)} \left\{ \varphi^{(0)}  ({\bm c}) + \frac{1}{2} \frac{\partial}{\partial {\bm c}} \cdot \left[ {\bm c} \varphi^{(0)} ({\bm c}) \right] \right\} \simeq \epsilon^{q} \sigma^{1-d} Q_{x}^{(1)} J^{(0)}
[{\bm c}| \varphi^{(0)}].
\end{equation}
If $Q_{x}^{(1)} \neq 0$, balance of this equation requires that $q=2$. Then multiplication of the equation by $c_{x}^{2}$ and integration over ${\bm c}$ yields
\begin{equation}
\label{4.12}
\gamma^{(0)} Q_{xx}^{(0)} =- 2 Q_{x}^{(1)} \sigma^{1-d} \int d{\bm c}\, c_{x}^{2} J^{(0)}[{\bm c}| \varphi^{(0)}],
\end{equation}
with
\begin{equation}
\label{4.13}
Q_{xx}^{(0)} = \int d{\bm c}\, c_{x}^{3} \varphi^{(0)}({\bm c}).
\end{equation}
Similarly, multiplication by ${\bm c}_{\perp}^{2}$ and integration over ${\bm c}$ gives
\begin{equation}
\label{4.14}
\gamma^{(0)} Q_{x \perp}^{(0)} = - 2 Q_{x}^{(1)} \sigma^{1-d} \int d{\bm c}\, c_{\perp}^{2} J^{(0)}[{\bm c}| \varphi^{(0)}],
\end{equation}
where  $Q_{x \perp}^{(0)}= Q_{x}^{(0)}-Q_{xx}^{(0)}$. Summation of Eqs. (\ref{4.13}) and (\ref{4.14}), taking into account that the collision operator $J^{(0)}$ conserves the kinetic energy, leads to Eq.\ (\ref{4.9}) providing a consistency test for the existence of the singular solution. Moreover, since the singular solution differs from the
gaussian, there is no reason to expect that the right hand side of Eqs.\ (\ref{4.12}) and (\ref{4.14}) vanish.
In summary, the asymptotic behavior of the solutions of the Boltzmann equation is qualitatively consistent with the existence of the ``singular'' solution discussed above. Going to higher order terms trying to actually determine the
asymptotic values of the moments of the singular solutions seems a formidable task.

\section{Molecular Dynamics simulation results}
\label{s5}
The Fourier state described in the previous sections is a ``bulk state'', in the sense that it is
expected to be shown by a dilute granular gas far away from the boundaries under the appropriate conditions, e.g. stationarity, no macroscopic mass flow, and gradients in only one direction \cite{GZyB97,BRyM00}. Whether or not this scenario can be generated with enough accuracy in experiments with real boundary conditions, or even in particle simulations with idealized ones, is not at all trivial. In principle, it could be thought that the bulk state would be reached by increasing the size of the system, while suitably scaling its properties
\cite{Ba75}. Nevertheless, due to the coupling between the gradients, the pressure, and the inelasticity in the Fourier state, the identification of the right way of scaling the system turns out to be far from trivial.
Although several different attempts were made trying to reach the Fourier state corresponding to the singular solution in the simulations, we did not succeed, possible because it is highly unstable. Consequently, attention will be restricted in the following to the regular solution, and all the references to theoretical predictions in the remaining of this section must be understood as dealing with that solution, although no established explicitly for the sake of brevity.

The macroscopic one-dimensional state we are considering is known to exhibit a hydrodynamic instability, leading to the development of transversal inhomogeneities \cite{LMyS02,BRMyG02}. In a two-dimensional system with periodic boundary conditions in the $y$ direction, the instability is controlled by the aspect ratio of the system $L_{y}/L_{x}$, where $L_{x}$ and $L_{y}$ are the dimensions of the system in the direction of the gradients and perpendicular to it, respectively. For given values of all the other parameters, the instability shows up when the aspect ratio exceeds a given critical value. Of course, it is possible that other instabilities occur in different regions of the space defined by the physical parameters of the system.

To investigate the validity of the theoretical predictions presented in the previous sections and its accessibility, molecular dynamics (MD) simulations of a system of inelastic hard disks ($d=2$) have been performed. The simulations started with the particles uniformly distributed on a square lattice and with a Gaussian velocity distribution. Several  ways of injecting energy through the walls were investigated, and it was concluded that the most efficient one to generate a bulk region in the system, are the so-called thermal walls \cite{Ce69,DyvB77}. In these walls, particles are absorbed at the surface and  instantaneously reemitted with a velocity determined by an equilibrium Maxwell-Boltzmann distribution, with a given wall temperature. The simulation data reported in the following have been obtained with two thermal walls located at $x=0$ and $x=L_{x}$, respectively, and with periodic boundary conditions in the $y$ direction.

In all the cases to be reported, it was found that, after a transient period, the system reached a macroscopic steady state with gradients only in the $x$ direction and no flow field. The results presented below have been time averaged, once the system was in the steady state, and also over several independent trajectories.  The temperature parameters of the thermal walls were fixed as follows. First, the temperature $T_{0}$ of the wall located at $x=0$ was arbitrarily fixed. Then, the temperature of the
wall at $x=L_{x}$ was determined by using the theoretical prediction for the temperature profile, Eq.\ (\ref{2.10}). In this
calculation, it was taken into account that the temperature of the thermal wall at $x=0$ ($x=L_{x}$) is expected to correspond to the temperature of the gas extrapolated to $x= -\lambda (0)$ ($x=L_{x}+ \lambda (L_{x})$), where $\lambda (x)$ is the local mean free path.

\begin{figure}
\begin{center}
\includegraphics[scale=0.5,angle=0]{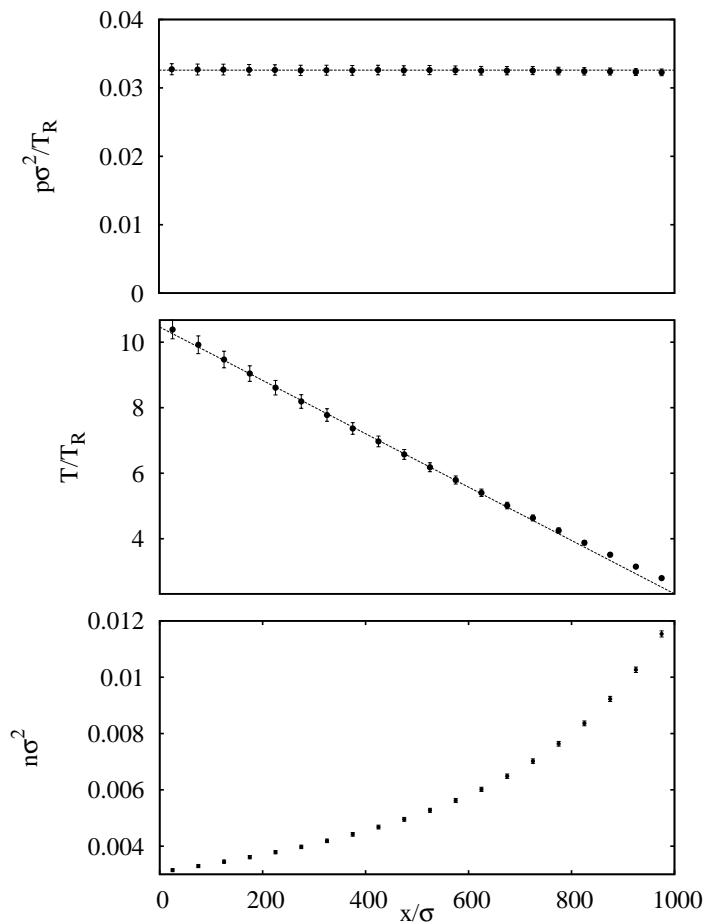}
\caption{Steady dimensionless pressure, temperature, and density profiles in a system of hard disks with $\alpha=0.99$. The temperature has been scaled with some arbitrary reference value, $T_{R}$, actually the initial temperature of the system. The symbols are simulation data, the dashed straight line in the pressure profile is a guide for the eye, and the dashed line in the temperature profile is a linear fit of the data in the interval $50 \sigma \leq x \leq 700 \sigma$. The simulation parameters are: $N=3 \times 10^{3}$, $L_{x}=10^{3} \sigma $, $L_{y}/L_{x}= 0.2$.  \label{fig4}}
\end{center}
\end{figure}

\begin{figure}
\begin{center}
\includegraphics[scale=0.5,angle=0]{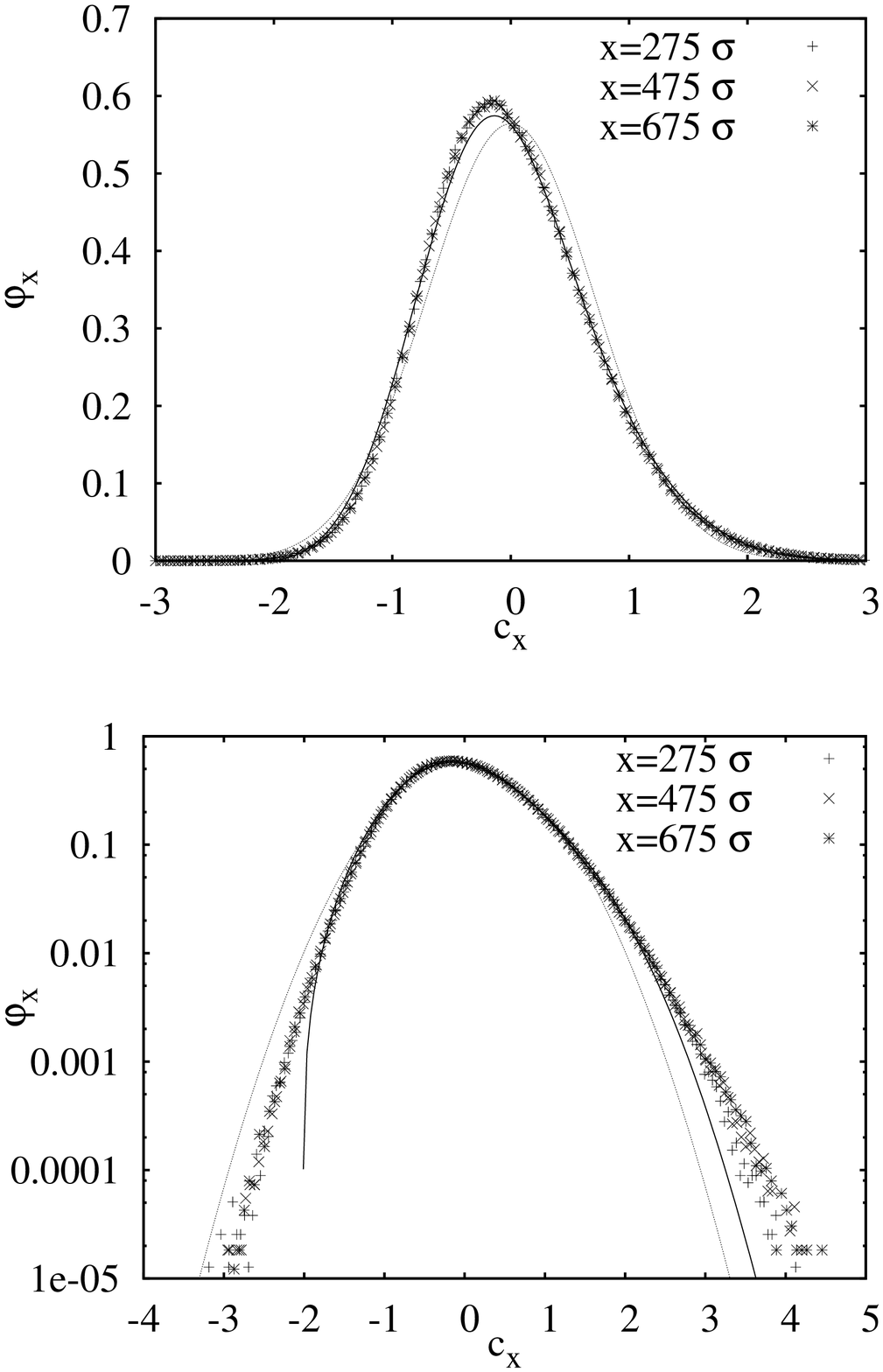}
\caption{The scaled distribution function for the same system as in Fig.\, \protect{\ref{fig2}}. The symbols correspond to MD simulation data at three different values of the coordinate $x$, as indicated in the insert. The solid line is the theoretical prediction in the first Sonine approximation derived in the text, Eq. (\protect{\ref{3.5a}}). The dotted line is the Gaussian, included as a reference.  \label{fig5}}
\end{center}
\end{figure}

Figure \ref{fig4} shows the pressure, temperature, and density profiles measured in
a system with $\alpha=0.99$. In agreement with the theoretical predictions, it is observed that the pressure is uniform and the temperature profile lineal, outside a boundary layer. In this case, the
bulk of the system, arbitrarily identified as the region in which the temperature profile is linear, extends over most of it.  A much more demanding and fundamental check of the theory is presented in
Fig. \ref{fig5}. There, the marginal velocity distribution $\varphi_{x}(c_{x})$ defined by
\begin{equation}
\label{5.1}
\varphi_{x}(c_{x}) = \int d c_{y}\,  \varphi ({\bm c})
\end{equation}
is plotted for three different values of $x$, namely for the layers centered at $x= 275 \sigma $, $475 \sigma$, and $675 \sigma$, respectively. The results for the other layers in the bulk are similar. Of course, in each layer the velocities have been scaled with the local temperature. It is seen that
the overlap of the data is very good over a wide range of velocities, hence confirming the
scaling of the distribution function in Eq.\ (\ref{2.8}). Moreover, the difference with a Gaussian (dotted line) is significant, while the theoretical prediction derived (solid line) here provides a much more accurate description of the distribution in the thermal velocity region, roughly $|c_{x}| < 2$ . To investigate the tails of the velocity distribution, the latter is also plotted on a logarithmic scale. As expected, the theoretical prediction derived here fails to correctly describe the tails, since it is based on the first Sonine approximation. On the other hand, the scaling seems to be verified within the statistical uncertainties.

Finally, in Fig. \ref{fig6}, the dimensionless heat flux defined in Eq.\ (\ref{3.7}) is plotted as a function of the position of the layer considered. Although for this property the statistical error, defined as the standard deviation of the measured values, is rather large, a quite uniform value along the system is neatly observed. Moreover, this value is in good agreement with the theoretical prediction
following from Eqs.  (\ref{3.8}) and (\ref{3.9}), with the coefficients determined by Eqs.\ (\ref{3.13}), (\ref{ap.1}), and (\ref{ap.2}).

{\begin{figure}
\begin{center}
\includegraphics[scale=0.7,angle=0]{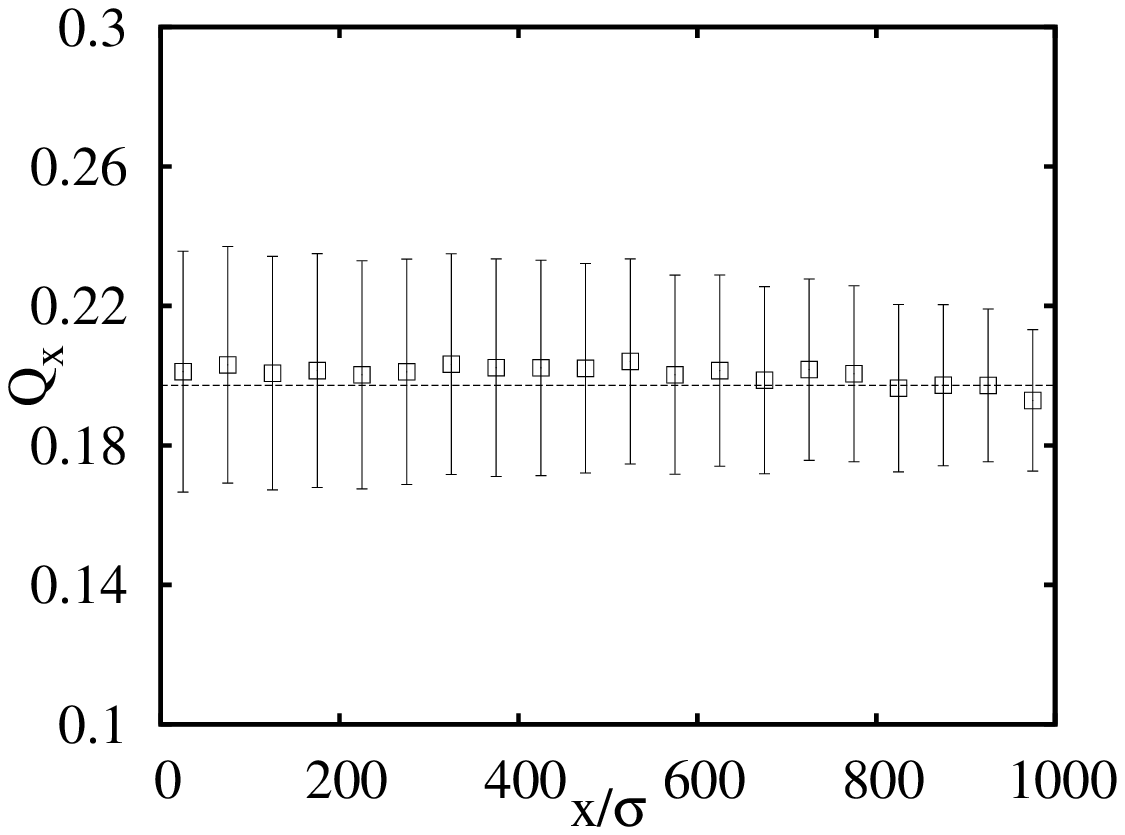}
\caption{The dimensionless heat flux $Q_{x}$ defined in Eq.\ (\protect{\ref{3.7}}), as a function of the reduced coordinate $x/\sigma$ for the same system as in Fig.\ \protect{\ref{fig1}}. The symbols are MD simulation results and the dashed line indicates the theoretical prediction obtained in this paper. \label{fig6}}
\end{center}
\end{figure}

Similar results have been obtained for other values of the coefficient of restitution in the interval $ 0.9 \leq \alpha < 1$, although the bulk region in which the Fourier state shows up becomes narrower as $\alpha$ decreases. In addition, the first Sonine approximation becomes less accurate. Nevertheless, there is strong evidence that the assumed Fourier state is present. To illustrate these comments, in Fig. \ref{fig7}, the temperature and heat flux profiles are plotted for a system with $\alpha = 0.9$. Now the bulk of the system, identified as the region in which the
temperature profile is linear and the heat flux uniform, is restricted to an interval roughly between $ 250 \sigma$ and $ 550 \sigma$. The reduced distribution function $\varphi ({\bm c})$ for three different positions inside the bulk region is shown in Fig.\ \ref{fig8}. Significant  discrepancies with the theoretical prediction in the first Sonine approximation are observed. On the other hand, there seems to be clear indication that the scaling assumed in Eq.\ (\ref{2.8}) holds in the bulk of the system. This scaling and the presence of gradients only in the direction perpendicular to the walls directly leads to the Fourier state discussed in Sec.\ \ref{s2}. A different issue is to solve the Boltzmann equation to obtain the explicit expression for the one-particle velocity distribution of the state. The first Sonine approximation developed here clearly fails to give an accurate description of the tails of the velocity distribution.

\begin{figure}
\begin{center}
\includegraphics[scale=0.5,angle=0]{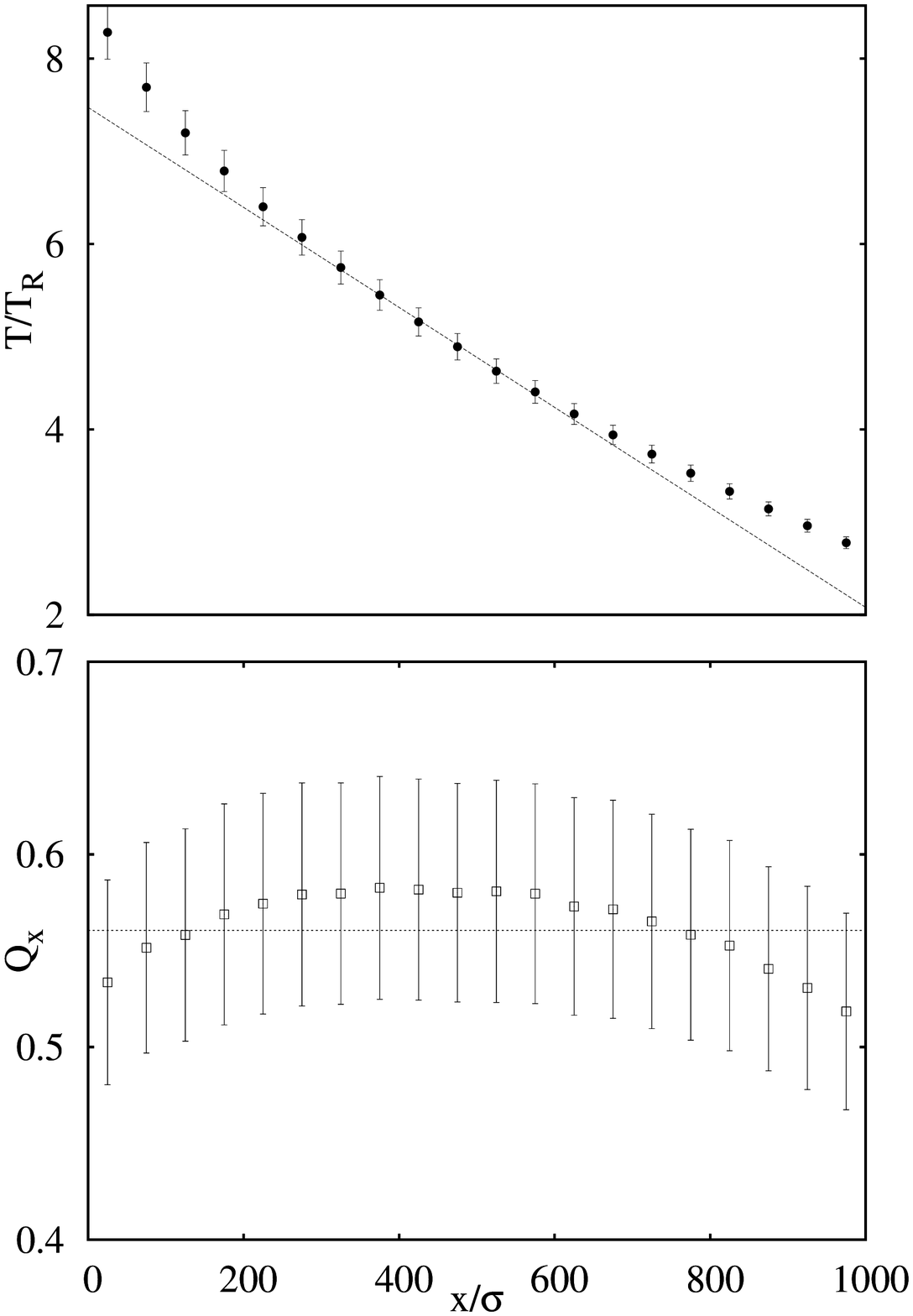}
\caption{Steady dimensionless temperature and heat flux profiles in a system of hard disks with $\alpha=0.9$. The temperature $T_{R}$ is the initial temperature of the system. The symbols are simulation data, the dashed line in the temperature profile is a linear fit of the data in the interval $250 \sigma- 550 \sigma$, and the dashed line in the heat flux profile is the theoretical prediction
in this paper. The simulation parameters are: $N=3 \times 10^{3}$, $L_{x}=10^{3} \sigma $, $L_{y}/L_{x}= 2$  .\label{fig7}}
\end{center}
\end{figure}

\begin{figure}
\begin{center}
\includegraphics[scale=0.7,angle=0]{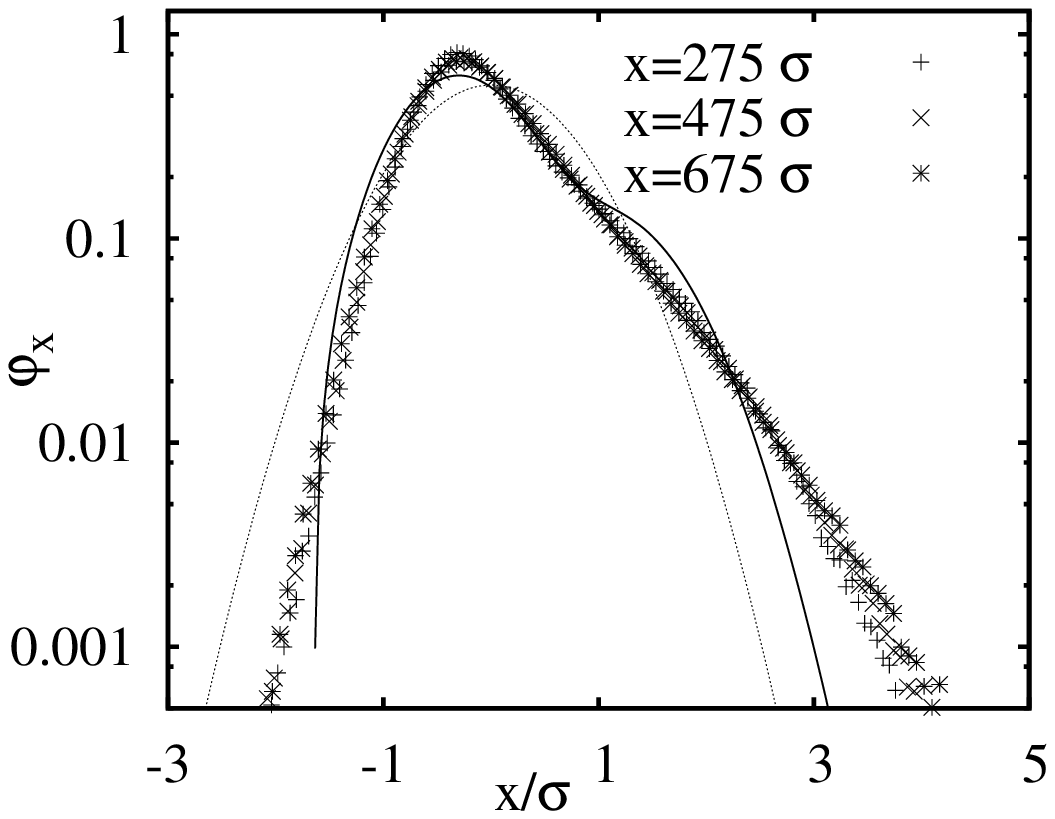}
\caption{The scaled distribution function for the same system as in Fig.\, \protect{\ref{fig5}}. The symbols correspond to MD simulation data at three different values of the coordinate $x$, as indicated in the insert. The solid line is the theoretical prediction in the first Sonine approximation derived in the text, Eq. (\protect{\ref{3.5a}}). The dotted line is the Gaussian, included as a reference.  \label{fig8}}
\end{center}
\end{figure}

By construction, the Sonine approximation is expected to lead to much more accurate results for the low velocity moments than for the complete one-particle distribution function itself. Moreover these moments contain the most relevant physical information on the macroscopic properties of the system.  A striking property of the Fourier state investigated here is the relationship (\ref{2.13}) between the temperature gradient, the pressure and the coefficient of restitution. Then, in Fig.\ \ref{fig9}, the MD simulation values obtained for the quantity $I \equiv \theta/ p \sigma$ in the bulk of the system has been plotted as a function of the restitution coefficient for the interval $ 0.9 \leq \alpha < 1$. These simulation values are seen to be in very good agreement with the theoretical prediction given by Eq.\ (\ref{3.12}), where the coefficients $a_{01}$, $b_{01}$, and $b_{10}$  are given by the solution of the system of equations (\ref{3.13}), (\ref{ap.1}) and (\ref{ap.2}) corresponding to the regular Fourier state. Also plotted in Fig.\ \ref{fig9} is the quantity $I$, as computed from the hydrodynamic Navier-Stokes equations derived from the Boltzmann equation by means of the Chapman-Enskog procedure in the first Sonine approximation \cite{BDKyS98,ByC01}. These equations also admit a solution with the
properties of the Fourier state \cite{BRyM00}. The Navier-Stokes prediction is seen to be very close to the result derived here. In fact, the difference between both predictions is smaller than the statistical uncertainties of the simulation data.

\begin{figure}
\begin{center}
\includegraphics[scale=0.7,angle=0]{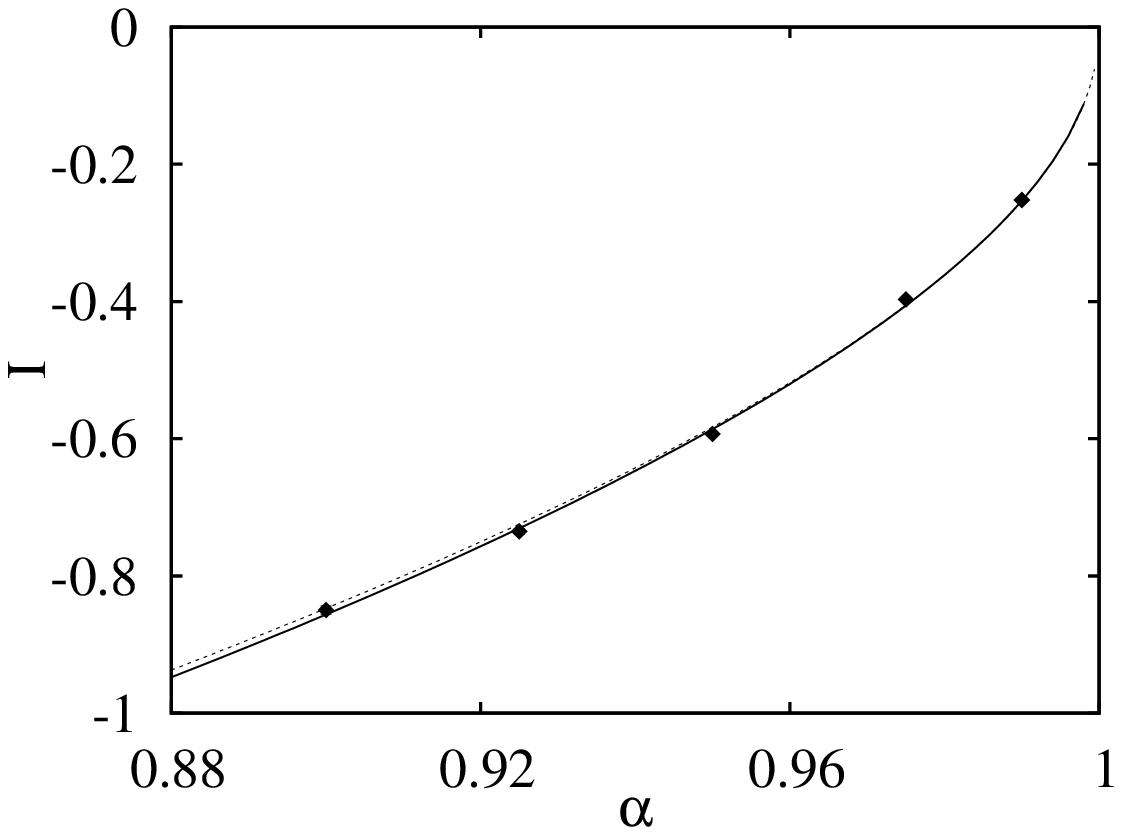}
\caption{Dimensionless ratio $I \equiv \theta/ p \sigma$ between the temperature gradient $\theta$ and the pressure $p$ in the Fourier state, as a function of the restitution coefficient $\alpha$, for a system of inelastic hard disks. The symbols are from MD simulations, the solid line is the theoretical prediction derived in this paper, and the dashed line is the result obtained from the inelastic Navier-Stokes equations.    \label{fig9}}
\end{center}
\end{figure}

Of course, another physically relevant moment is the heat flux $q_{x}$, or its dimensionless form $Q_{x}$ defined in Eq. (\ref{3.7}). This quantity is shown as a function of the restitution
coefficient in Fig. \ref{fig10}. The observed qualitative behavior is easily understood, since it follows from Fig. \ref{fig7} that the magnitude of the temperature gradient increases as $\alpha$ decreases. Again, there is a quite  good agreement between the MD simulation results and the expression found in this paper, i.e. Eqs. (\ref{3.7})-(\ref{3.9}) and $b_{01}$ and $b_{10}$ obtained as indicated above and discussed in Sec.\ \ref{s3} for the regular solution. As in Fig.\ \ref{fig9}, the result obtained from the hydrodynamic  Navier-Stokes equations, in the first Sonine approximation, for the Fourier state has also been included, and it differs very slightly from the solution derived here.

\begin{figure}
\begin{center}
\includegraphics[scale=0.7,angle=0]{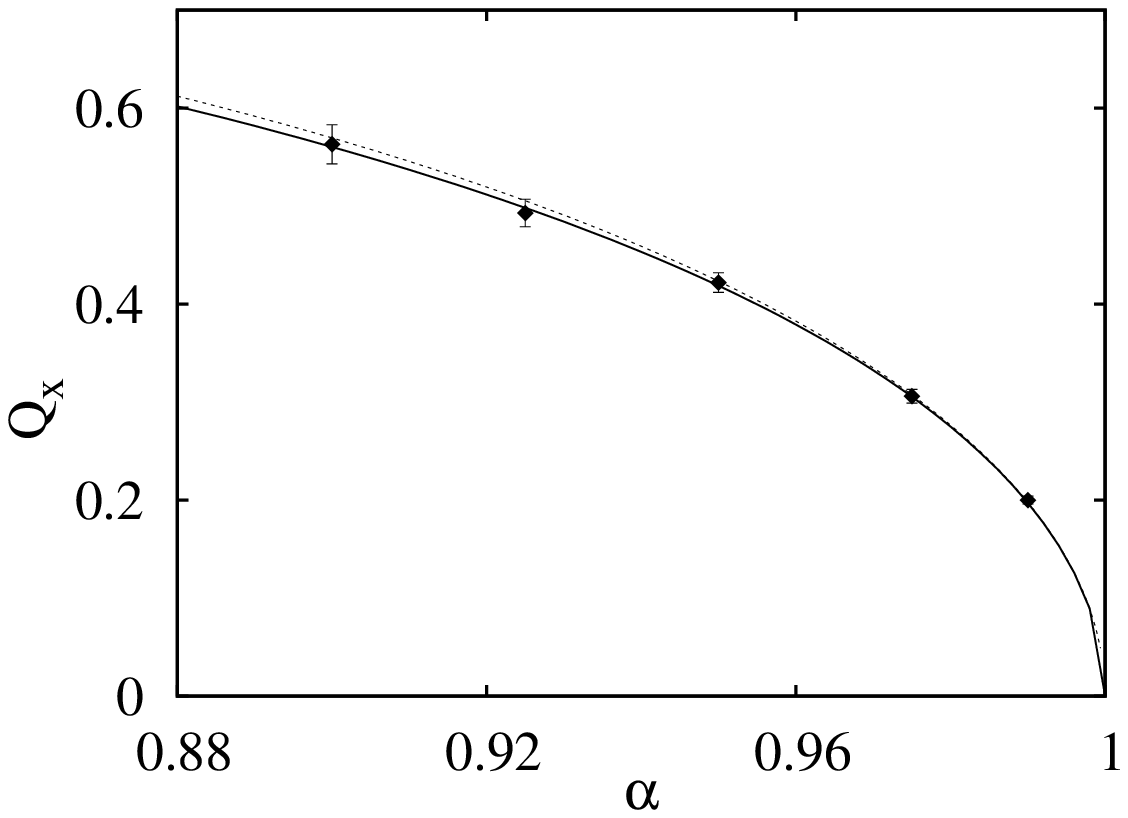}
\caption{Dimensionless heat flux $Q_{x}$ in the Fourier state as a function of the restitution coefficient, for a system of inelastic hard disks. The symbols are from MD simulations, the solid line is the theoretical prediction derived in this paper, and the dashed line is the result obtained from the inelastic Navier-Stokes equations.    \label{fig10}}
\end{center}
\end{figure}

\section{Summary and discussion}
\label{s6}
Two non-trivial solutions of the Boltzmann equation for smooth inelastic hard spheres or disks have been
investigated. They describe a stationary state with constant pressure, temperature gradient in one  direction, and no macroscopic mass flow. Quite peculiarly, the temperature profile is strictly linear and the
heat flux can be expressed as proportional to it, without nonlinear  contributions. In other words, the macroscopic  state does not
present any kind of rheological hydrodynamic effects. For this reason the state was termed Fourier state. Although no rigorous mathematical proof of the existence of the solutions is given, it has been shown that
supposing they exist, a consistent expression for them can be found under well defined approximations.
One of the solutions is singular in the sense that it does not reduce to any solution of the elastic Boltzmann equation when the limit of the coefficient of normal restitution going to unity is considered. It provides an example of a possible normal solution of the Boltzmann equation that is not captured by the Chapman-Enskog procedure to solve it. On the other hand, the other solution, termed regular, tends to an equilibrium Gaussian in the elastic  limit. It is worth to stress that although both distribution functions are quite different, they lead to qualitatively similar macroscopic states.

The singular state was never observed in the molecular dynamics simulations we carried out,  while the theoretical predictions corresponding to the other solution are in good agreement with the simulation data, in the parameter region in which the Fourier state is accesible to the simulations, e.g. not too strong inelasticity.

In this study, attention has been restricted to the first Sonine approximation, mainly for the sake of simplicity and also in order to derive results in an analytical form. From this starting point, two extensions are possible. The first one is to incorporate more terms in the Sonine expansion of the distribution function, solving the resulting equations numerically. A way of implementing this is by following the method developed in \cite{NBSyG07}, that is tailored to solve the Boltzmann equation without considering any gradient expansion. The second extension refers to the tails of the distribution function. An asymptotic analysis of the inelastic Boltzmann equation for the Fourier state shows that for large values of the component of the velocity in the opposite direction to the increase of the temperature, the marginal velocity distribution exhibits an algebraic tail \cite{BCMyR01}. This feature can also be analyzed in an numerically exact way by using the method presented in ref. \cite{NBSyG07}. This would allow, for instance, to predict the amplitude of the algebraic decay.

Future work should also attempt to put the existence of the Fourier state solution(s) of the Boltzmann equation on a more rigorous mathematical basis.  Even if  only the regular solution exists and this happens in some well defined limit,
it should be an important result, since it would provide an explicit non-trivial normal solution of the Boltzmann equation. Moreover, the Fourier state also offers the opportunity of studying hydrodynamic fluctuations and correlations in a far from equilibrium state, without resorting to expansions in the gradients of the fields.

\ack

This research was supported by the Ministerio de Educaci\'{o}n y
Ciencia (Spain) through Grant No. FIS2008-01339 (partially financed
by FEDER funds).

\appendix

\section{Two moment equations for the Fourier state in the first Sonine approximation}
\label{ap1}

Multiplication of Eq. (\ref{2.15}) by $c_{x}^{3}$ and integration over ${\bm c}$ leads to
\begin{eqnarray}
\label{ap.1}
&& 64 d (d+2)(d+4)(1-\alpha) \left[ 1 + 2 (d-1)a_{01} \right] \nonumber \\
&&  - (d-1) \left[ 13 + 51 \alpha - 15 d -33 \alpha d -4d^{2}(1+ 3 \alpha ) \right] b_{01}^{2}
\nonumber \\
&& -9 \left[ 75+4d(21+4d-9 \alpha)-139 \alpha \right] b_{10}^{2} \nonumber \\
&& - 6 (d-1) \left[ 29 +8d (5 + d - 3 \alpha) - 93 \alpha \right] b_{01} b_{10}  =  0.
\end{eqnarray}
In a similar way, multiplication by $c^{2}c_{x}$ yields
\begin{eqnarray}
\label{ap.2}
&& 64 d (d+2) (d+4) (1-\alpha) \left[d+2+(d+1)(d+4)a_{01} \right] \nonumber \\
&& +(d-1) \left[ 262 -390 \alpha - 7 d + 231 \alpha d  - d^{2} \left( 193 + 44d - 3(43
+4d) \alpha \right) \right] b_{01}^{2} \nonumber \\
&& - 9 \left[ 246+235 d +44 d^{2} - (22+3d)(17+4d) \alpha \right] b_{10}^{2}
\nonumber \\
&& - 6 (d-1) \left[ 250 - 378 \alpha + 237 d +44 d^{2} - 3d(47+4d)\alpha \right] b_{01} b_{10} =0.
\end{eqnarray}

In the calculations leading to the above relations, Eq. (\ref{3.12}) has been used and only terms
up to second degree in $a_{01}$, $b_{01}$, and $b_{10}$ have been kept as discussed in Sec \ref{s3}.
The circles plotted in Fig.\ \ref{fig1} have been obtained by solving Eqs. (\ref{3.13}), (\ref{ap.1}), and (\ref{ap.2}) for $d=2$.

\end{document}